\documentclass[prl,showpacs,twocolumn,preprintnumbers]{revtex4}

\usepackage{graphicx}
\usepackage{epsfig}
\usepackage{dcolumn}
\usepackage{bm}

\begin{document}

\title{Spin readout and initialization in a semiconductor quantum dot}
\author{Mark Friesen}
\email{friesen@cae.wisc.edu}
\author{Charles Tahan}
\author{Robert Joynt}
\author{M.~A.~Eriksson}
\affiliation{Department of Physics, University of Wisconsin, Madison, Wisconsin 53706, USA}

\begin{abstract}
Electron spin qubits in semiconductors are attractive from the viewpoint of
long coherence times. However, single spin measurement is challenging.
Several promising schemes incorporate ancillary tunnel couplings that may
provide unwanted channels for decoherence. Here, we propose a novel
spin-charge transduction scheme, converting spin information to orbital
information within a single quantum dot by microwave excitation. The same
quantum dot can be used for rapid initialization, gating, and readout. We
present detailed modeling of such a device in silicon to confirm its
feasibility.
\end{abstract}

\maketitle

Quantum dot qubits lie at the point of intersection between quantum
computing and few-electron spintronics. The long coherence times and novel
quantum mechanical properties of electron spins, the attraction of
scalability, and the shared technology base with the digital electronics
industry fuel a large research effort to develop these systems. Several
proposals have been put forward for spin qubits in semiconductors \cite%
{loss98,kane98,vrijen,levy,friesen03}. In particular, exchange coupled gates
in semiconductor dots appear to be workable \cite{friesen03}, and
several components of qubit technology have been demonstrated recently \cite%
{elzerman}. However, the combined challenge of preparing, storing and
measuring spins is formidable.

Measurements of spin qubits pose a special challenge. On the one hand,
qubits should be well isolated from their environment to avoid decoherence.
On the other hand, it is necessary to individually couple the qubits to an
external measurement device. Qubit initialization involves an additional
dissipative coupling to the environment. For quantum computing, we must
initiate such couplings selectively, and with sufficient strength to perform
the operations quickly. Indeed, scalable quantum computing relies on
fault-tolerant quantum error correction algorithms, involving frequent,
parallel measurements, and a steady supply of initialized qubits \cite{Sho96}%
. Fortunately, rapid and sensitive quantum measurement techniques involving
radio frequency single electron transistors (rf-SETs) have been developed 
\cite{scholkopf}. rf-SETs have been used to detect the tunneling of
individual electrons in semiconductor devices \cite{rimberg03}.

We have previously designed a quantum dot architecture specifically for the
purpose of manipulating electron spins for fast and accurate two-qubit
operations that serve as universal gates for quantum computations \cite%
{friesen03}. Recent experimental results have shown that decoherence does
not pose a fundamental problem for such gate operations \cite{lyon03}. \
Using special qubit geometries \cite{friesen02}, it should be possible to
perform reliable gate operations in silicon quantum dots at rates between
about 1~MHz and 1~GHz \cite{note}. The question is now whether similar
speeds and reliable operation can be achieved for measurement and
initialization operations. In this paper, we show how to extend our
architecture to enable the readout and initialization of the state of a
single spin. The general scheme does not assume a specific material
system.  However, in this Letter we focus upon silicon-based devices 
because of their desirable coherence properties. 

Our method relies on the idea of converting spin information to charge
information discussed by Loss and DiVincenzo \cite{loss98}, and using single
electron transistors to read out the resulting spin state as proposed by
Kane \textit{et al.} \cite{kane98}, who posit spin-dependent charge motion onto
impurities in Si. Martin \textit{et al.} \cite{martin} have proposed a
scheme for single spin readout that also converts spin information into
charge information in an electron trap near a conducting channel. The
resistance of the channel depends on the occupation of the trap, which in
turn can be made to depend on the spin. \ Our proposed device has the
additional feature in that it enables both readout \textit{and} rapid
initialization of the spin state. Rapid initialization (as opposed to
initialization by thermalization) is just as essential as readout, and just
as difficult to carry out. \ Our design is simple in that it obviates the
need for spin-polarized leads or ancillary qubits. \ We go beyond conceptual
design--the full three-dimensional system is simulated self-consistently and
the operating parameters are thereby optimized. \ This enables us to
demonstrate quantitatively that the device is in a working regime.

The basic Si-SiGe heterostructure is described in Ref. \cite{friesen03}. The
main point for present purposes is that the active layer is pure strained
Si, which minimizes decoherence from spin-phonon coupling \cite{tahan02}.
The qubits are gated quantum dots which hold one electron, and the gate
geometry is the key to our scheme. \ Specifically, the electrons are
confined in asymmetric lateral wells, such that orbital excitation results
in lateral center-of-charge movement. \ Figure~1 shows the first two orbital
states of an electron confined to an asymmetric box, with center-of-charge
positions varying by a distance $\Delta y$. If an external radiation source of
frequency $E_{12}/h$ drives the system between the ground and first excited
states, $n=1$ and $n=2$, the center-of-charge of the electron will oscillate
in time at the Rabi frequency $\nu _{R}$. This motion can be detected by a
sensitive electrometer, such as a single electron transistor or a quantum
point contact. With the system placed in a magnetic field, charge motion can
be generated in a \emph{spin-dependent} fashion. We define the logical 0 and
1 of the qubit as the spin up and spin down states of the electron in its
orbital ground state, and use the orbital excited state only during
initialization and readout. \ The system is now driven at the readout
frequency $\nu_{12}(B)=E_{12}/h-g\mu_{B}B/h$. This causes charge motion only
if the qubit was in the down state at the time of measurement. Although this
transition is forbidden at the electric dipole level because of the spin
flip, spin-orbit coupling allows for a nonzero transition rate, as described
below.

\begin{figure}[tbp]
\centerline{\epsfxsize=2.8in \epsfbox{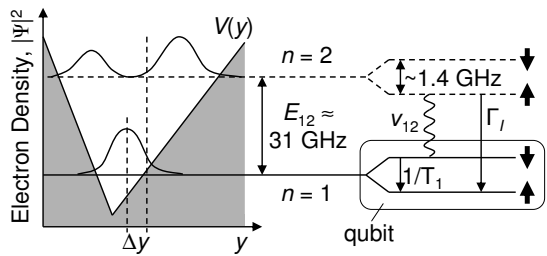}} 
\caption{ Schematic example of a one-dimensional, asymmetric confining
potential. Quantum dot energy states and transition rates for readout and
initialization are shown.}
\label{fig:scheme}
\end{figure}
Crucially, this architecture also allows for rapid initialization of the
qubit. \ Consider exposing a random ensemble of qubits to radiation of
frequency $\nu _{12}$. \ An electron in the $(1,\downarrow )$ state will be
excited to the level $(2,\uparrow )$ and experience a relatively fast
relaxation to the ground state $(1,\uparrow )$, compared to a spin-flip
relaxation to the level $(1,\downarrow )$. \ The net result is to rapidly
polarize and thereby initialize the qubits. By varying the gate voltages (and
thus $\nu _{12}$) on individual dots we ensure that only desired qubits are
brought into resonance. Clearly, understanding the competition between the
three time scales, $1/\nu _{R}$, $1/\Gamma _{I}$, and $T_{1}$, as a function
of material parameters and gate potentials is the key to utilizing this
device for readout and initialization. \ Note that $T_{1}$ represents a
thermal initialization time from $(1,\downarrow )$ to $(1,\uparrow )$. Since 
$T_{1}$ is a decoherence time, we require $1/\nu _{R}\ll T_{1}$. The Rabi
oscillation frequency $\nu _{R}$ depends on the incident intensity and is
therefore controllable, within limits. Robust measurement requires that many
Rabi oscillations occur before orbital decay: $1/\nu_R\lesssim 1/\Gamma_I$.

In order to understand the competition of time scales, we introduce an rms
interaction energy, time averaged (as indicated by \{\}) over an optical
cycle, which expresses the strength of the interaction in the electric
dipole approximation. \ This defines the Rabi oscillation frequency \cite%
{shore}: $|h\nu_{R}|^{2}=\{|V^{E1}|^{2}\}$. Here, $V^{E1}=(-e\hbar
E_{0}/m^{\ast }E_{12})\hat{\epsilon}\cdot \mathbf{p}$ is the dipole term in
the Hamiltonian, $|E_{0}|^{2}$ is twice the mean value of $|\mathbf{E}%
(t)|^{2}$ averaged in time, and $\hat{\epsilon}$ is the polarization unit
vector. The electric field $E_{0}$ inside the semiconductor with dielectric
constant $\varepsilon =\varepsilon_r \varepsilon_0$ 
is related to the intensity of the external
radiation $I$ by $E_{0}=\sqrt{2I/c\varepsilon _{0}\sqrt{\varepsilon_r}}$.

The dipole Hamiltonian does not flip the spin directly, but spin-orbit
coupling causes each qubit state to be an additive mixture of up and down
spin. 
In the 2D limit, the spin-orbit (so) Hamiltonian is dominated by the bulk
[Dresselhaus (D)] and structural [Rashba (R)] inversion asymmetry terms, $%
H_{\rm so}=H_{D}+H_{R}$, where $H_{D}=\beta (p_{y}\sigma _{y}-p_{x}\sigma _{x})$
and $H_{R}=\alpha (p_{x}\sigma _{y}-p_{y}\sigma _{x})$. 
Including $H_{so}$ perturbatively gives a nonzero
dipole matrix element, and for light polarized in the $y$~direction, the
readout frequency is given by 
\begin{equation}
\left\vert \nu _{R}\right\vert \simeq \frac{eE_{0}}{2\pi h\nu _{12}(B)}\sqrt{%
\alpha ^{2}+\beta ^{2}}|\langle 2|y\partial _{y}|1\rangle |.
\label{eq:readout}
\end{equation}%
Note that the applied radiation need not be circularly polarized for this
readout scheme. The Dresselhaus and Rashba parameters $\alpha $ and $\beta $
depend on intrinsic material properties, device design, and external
electric field.  For GaAs, both theoretical and experimental values
vary widely \cite{silva}: 
$\alpha =1-1000$ m/s and $\beta =1000-3000$ m/s. In a centrosymmetric crystal
like silicon which has no bulk inversion asymmetry, $\beta =0$. The one
known data point for a SiGe two-dimensional electron gas gives $\alpha
\simeq 8$~m/s \cite{alphainsi-rossler}, which we use in our estimates below.

The relaxation of our quantum dot to its ground state enables spin
polarization, but limits or even inhibits readout if it occurs too quickly.
This problem has been addressed by Khaetskii and Nazarov in GaAs quantum
dots \cite{khaetskii}.
In silicon, where there is no piezoelectric interaction, we calculate the
relaxation rate via the Golden Rule with the usual deformation potential
electron-phonon Hamiltonian \cite{ridley}. At sufficiently low temperatures $%
(T<1K)$, optical polar phonons and multiphonon processes do not contribute.
By considering only longitudinal phonons, with dispersion $\omega =v_{l}q$,
and using the long-wavelength approximation $e^{i\mathbf{k\cdot r}}\simeq 1+i%
\mathbf{k\cdot r}$, we obtain the orbital decay rate due to electron-lattice
coupling: 
\begin{equation}
\Gamma _{I}=\frac{(E_{12})^{5}}{6\pi \hbar ^{6}v_{l}^{7}\rho}\left( \Xi
_{d}+\Xi _{u}/3\right) ^{2}\sum_{i}|\langle 1|x_{i}|2\rangle |^{2},
\label{eq:relaxation}
\end{equation}%
where $\rho $ is the mass density, and $\Xi _{d}$ and $\Xi _{u}$ are the
deformation constants. In strained systems, transverse phonons can also be
important, as we shall discuss in a future publication \cite{friesen04}.

\begin{figure}[tbp]
\vskip .6in
\centerline{\epsfxsize=2.5in \epsfbox{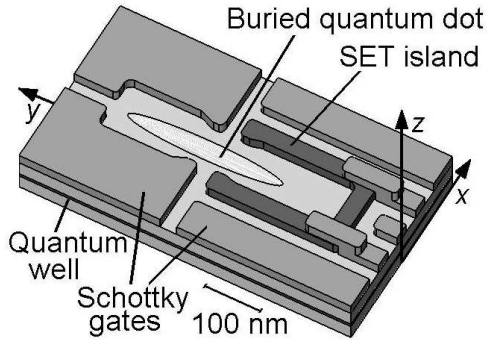}} 
\caption{ The readout device studied in Figs.~3 and 4. Many similar devices
were modeled in this work.}
\label{fig:device}
\end{figure}

We perform a numerical analysis to obtain performance characteristics for
the proposed measurement scheme. \ The numerical techniques used are an
extension of those used in Refs.~[\onlinecite{friesen03,friesen02}]. The gate potentials, the electronic orbitals, and
their corresponding image potentials (arising predominantly from the
metallic gates) are computed self-consistently by a combination of
three-dimensional finite element and diagonalization techniques.
Specifically, we determine the readout oscillation frequency Eq.~(\ref%
{eq:readout}), the orbital decay rate Eq.~(\ref{eq:relaxation}), and the
coupling sensitivity of the qubit electron to an integrated SET. \ We have
modeled numerous devices and a full account will be published elsewhere \cite%
{friesen04}. \ One of the most promising is shown in Fig.~\ref{fig:device}.
This is a 6~nm strained silicon quantum well sandwiched between layers of
strain-relaxed silicon-germanium (Si$_{85}$Ge$_{15}$). The bottom barrier
(30~nm) separates the quantum well from a grounded metallic back-gate. The
top barrier (30~nm) separates the quantum well from lithographically
patterned Schottky top-gates, whose voltages can be controlled
independently. We consider negative gate potentials, which provide lateral
confinement of the quantum dot through electrostatic repulsion. The main
features of the quantum dot are that it is narrow, long, and slightly
asymmetric. The \textit{narrow} feature ensures that the excited orbitals
are non-degenerate, so a microwave field with narrow line width will not
induce unwanted transitions. The \textit{long} feature ensures that the
energy splitting $E_{12}$ (and thus $\Gamma_I$) will be small. The 
\textit{asymmetry} ensures that excited orbitals will provide charge motion.
Figure~3 shows the confinement potential for the device of Fig.~2; the 
asymmetry of the dot potential is apparent.

Our most interesting results are obtained in the regime where the gate and
image potentials are comparable in size. The effective confinement
potential, including images, is rather complicated due to the inhomogeneous
gate arrangement. Indeed, it is a very poor approximation to neglect
screening in this system. The computed wavefunctions for our dot are shown
in Fig.~3.

We have also incorporated a SET into our proposed readout device, as shown
in Fig.~\ref{fig:device}. The island of the SET is sheathed by a thin, 2~nm
layer of silicon dioxide, and tunnel-coupled to adjoining source and drain
gates. A third, capacitively-coupled gate is placed nearby, to provide full
control over both the charge and the potential of the island. When the
device is operated in the Coulomb blockade regime, it becomes a sensitive
electrometer \cite{schoelkopf98}. In the gate arrangement of Fig.~\ref%
{fig:device}, the SET plays two roles. First, voltage control of the island
allows us to vary the size of the dot and thus the energy splitting $E_{12}$%
. Second, capacitive coupling to the quantum dot enables detection of its
orbital state. The scheme works as as follows. Since the dot orbitals are
spatially distinct, they induce different amounts of charge on the SET
island. Consequently, transport currents through the SET will reflect the
orbital states of the dot. The device exhibits optimal sensitivity if biased
at the half-maximum of the conductance peak. The third SET gate has been
introduced specifically to adjust this working point. As expected for our
geometry, we find that the SET couples most strongly to the excited
electronic orbital.

\begin{figure}[tbp]
\centerline{\epsfxsize=2.5in \epsfbox{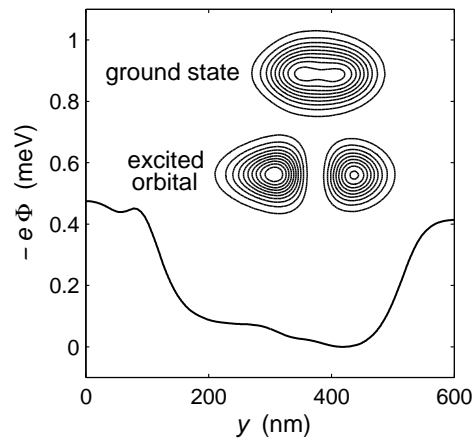}}
\caption{Electrostatic confinement potential and qubit electron
wavefunctions. The bare potential shown here is obtained in the quantum well
along the device symmetry line, $x=0$. The
contour plots show the electron probability densities in the $x$-$y$ plane. }
\label{fig:phipsi}
\end{figure}
For the device shown in Fig.~\ref{fig:device} we obtain an energy splitting
of $E_{12}=0.129~\mathrm{meV}=31.2~\mathrm{GHz}$ between the two lowest 
orbital states, and the
dominant matrix elements $|\langle 1|y|2\rangle |=48$~nm and $|\langle
2|y\partial_{y}|1\rangle |=3.6$. From these results we obtain the readout
oscillation frequency $\nu _{R}=5.5\times 10^{5}\sqrt{I}~\mathrm{Hz}$ (for
microwave intensity $I$ in units of W/m$^{2}$), and the orbital decay rate
for spontaneous phonon emission $\Gamma _{I}=12.7$~MHz. The orbital decay
rate for emission of a photon is very much less: 
$\Gamma_\gamma =e^2 E_{12}^3 |\langle 1 | y | 2 \rangle|^2 \sqrt{\varepsilon}
/3\pi \hbar^4c^3\varepsilon_0^{3/2}=6.5$~Hz, and
therefore is not a limiting factor in this scheme.  All results are
obtained at an ambient temperature of $T=100$~mK and a magnetic field of 
$H=0.05~\mathrm{T}=1.40~\mathrm{GHz}$.  The dielectric constant of Si is
$\varepsilon = 11.9\varepsilon_0$.  \ Note that the natural widths of the 
states are small compared
to their separation. \ The charge on the SET island was computed by integrating
$\mathbf{n}\cdot\mathbf{D}$ over the surface.  The excess or induced
charge of the excited quantum dot orbital, relative to its ground state, was
found to be $\Delta Q=0.052e$. 
The corresponding center-of-charge motion in the dot is $\Delta
y=4.3$~nm. However, the charge motion does not track closely with $\Delta Q$;
the latter is determined primarily by the capacitive coupling between the
dot and the SET. Finally, we find that by changing the bias voltage on the
SET by 20\% (thus reducing the dot size), the excitation resonance frequency 
$\nu_{12}$ changes by 8~GHz.

\begin{figure}[tbp]
\centerline{\epsfxsize=2.5in \epsfbox{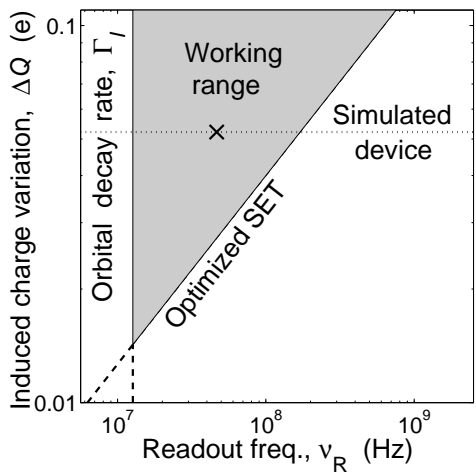}}
\caption{Summary of operation parameters for the device shown in Fig.~2.
The shaded region shows the operating range, with constraints set by the
sensitivity of the SET and the spontaneous decay rate of the upper readout
state. The dotted line shows the actual operating range of the device when
the intensity of the incoming radiation is left variable. A likely working point 
of $\protect\nu_R$, marked by $\times$,
corresponds to an intensity $I=7.2~\mathrm{nW}/\mu\mathrm{m}^2$. }
\label{fig:rfset}
\end{figure}
The prospects for initialization and readout in this scheme are partially
summarized in Fig.~4. \ Theoretical considerations of shot noise \cite%
{schoelkopf98,korotkov99} place an upper bound of about $4\times 10^{-6}e/%
\sqrt{\mathrm{Hz}}$ on the detection sensitivity for charge induced on the
island of an optimized rf-SET, as a function of the measurement bandwidth.
Similarly, the decay rate of the excited electronic orbital places a lower
bound on the readout oscillation frequency $\nu_{R}$. The latter is a
function of microwave intensity.
By directing $70$~pW of microwave power
onto a dot of size $\sim0.1\times0.1~\mu\mathrm{m}^2$,
as consistent with low-temperature transport experiments, we would obtain 
the $\nu_R$ working point marked in Fig.~4, with
$1/\nu_R\lesssim 1/\Gamma_I$.  Sample
heating outside the dot can be minimized by using integrated on-chip 
antennas to focus the power \cite{blick95}.  We must
also satisfy the requirement $T_s\gg 1/\Gamma_I=78.7~\mathrm{ns}$ where 
$T_{s}$ is the spin decoherence time. \ Theoretical estimates for $T_{1}$ in
Si quantum dots exceed 1~ms \cite{tahan02}. Experiments on dots have yet
to be performed, but spins of electrons in impurity donor states should have
similar lifetimes. \ Recent work on electron spins in $^{28}$Si:P 
\cite{lyon03} would imply that $T_{s}>\mbox{60 ms}$. \ Hence the
crucial hierarchy of timescales $1/\nu _{R}\lesssim1/\Gamma _{I}\ll$\ $T_{s}$ 
should be achievable.

In summary, we have proposed, simulated, and analyzed a scheme for measuring single
electron spins in a quantum dot via spin-charge transduction.  Adequate charge 
detection sensitivity is provided by an integrated rf-SET.  The scheme also
enables fast initialization.  

We thank Alex Rimberg and the quantum computing group at the University of
Wisconsin-Madison for many beneficial discussions. Our work was supported by
the U.S.\ Army Research Office through ARDA, and the National Science
Foundation through the QuBIC program, DMR-0079983, and DMR-0081039.


\end{document}